\documentclass[a4paper,20pt]{article}
    \usepackage[T1]{fontenc}
    \usepackage{graphicx}
    \usepackage{epsfig}
    \usepackage{amsmath}
    \usepackage{amsfonts}
    \renewcommand{\abstract}{}
\begin{document}
\makeatletter
\renewcommand{\@oddhead}{\textit{Advances in Astronomy and Space Physics} \hfil \textit{A.V. Tugay, S.L. Shihov}}
\renewcommand{\@evenfoot}{\hfil \thepage \hfil}
\renewcommand{\@oddfoot}{\hfil \thepage \hfil}
\fontsize{11}{11} \selectfont

\title{Surface Brightness Profiles of Seyfert Galaxies}
\author{\textsl{A.\,V.~Tugay, S.\,L.~Shihov}}
\date{}
\maketitle
\begin{center} {\small
Taras Shevchenko National University of Kyiv, Glushkova ave., 4, 03127, Kyiv, Ukraine\\
tugay.anatoliy@gmail.com}
\end{center}

\begin{abstract}
We built r-band surface brightness profiles by SDSS data for 16 Seyfert galaxies observed in Crimean Astrophysical Observatory. Obtained profiles can be used for finding more accurate lightcurves for these galaxies.
\end{abstract}

\section*{Introduction}
\indent \indent The study of active galactic nuclei (AGN's) is actual branch of modern astrophysics. AGNs have been observed at Crimean Astrophysical Observatory (CrAO) since 1970-th \cite{sergeev05}. One of the tasks of AGN study in CrAO is building optical lightcurves and comparing them with X-ray ones\cite{sergeev07}. The total list of 58 AGN's observed at CrAO was published in \cite{doroshenko05a},\cite{doroshenko05b},\cite{doroshenko07} and \cite{doroshenko08}. Those galaxies were observed on different telescopes with different construction of radiation recievers. In each observation when brightness of AGN is measured, observers take emission from central part of galaxy with some radius. That radius may differ in each observation. To unite such different data correctly we should find surface brightness profiles for studied galaxies. Then one could subtract magnitude from central region, obtain real brightness of point-like nucleus and build precisious lightcurve.
\section*{The method of calculations}

\indent \indent 
We used images from Sloan Digital Sky Survey (SDSS). We found SDSS data of 16 Seyfert galaxies from CrAO AGN sample. SDSS has 5 optical bands: u, g, r, i and z. As the galaxy surface brightness must be equal to $ 25^m/\square ''$ at its edge (by definition of galaxy size), we were interested of finding and using galaxy sizes to check or profiles. SDSS collaboration finded isophote sizes (major and minor axes) in r-band for large number of galaxies. That sizes are bounded by isophote of $ 25^m/\square ''$. So we used r-band images for obtaining surface brightness profiles. We calculated counts on images in rings of 1 arcsec width, so background inhomogeneties were averaged. We used ring-like background regions with minor radius equal to major semiaxis of galaxy and major radius depending of position of nearest bright source or end of plate. In the most cases that radius is equal to 120$''$ (see Table 1).

\begin{figure}[!h]
\centering
\begin{minipage}[t]{.45\linewidth}
\centering
\epsfig{file = 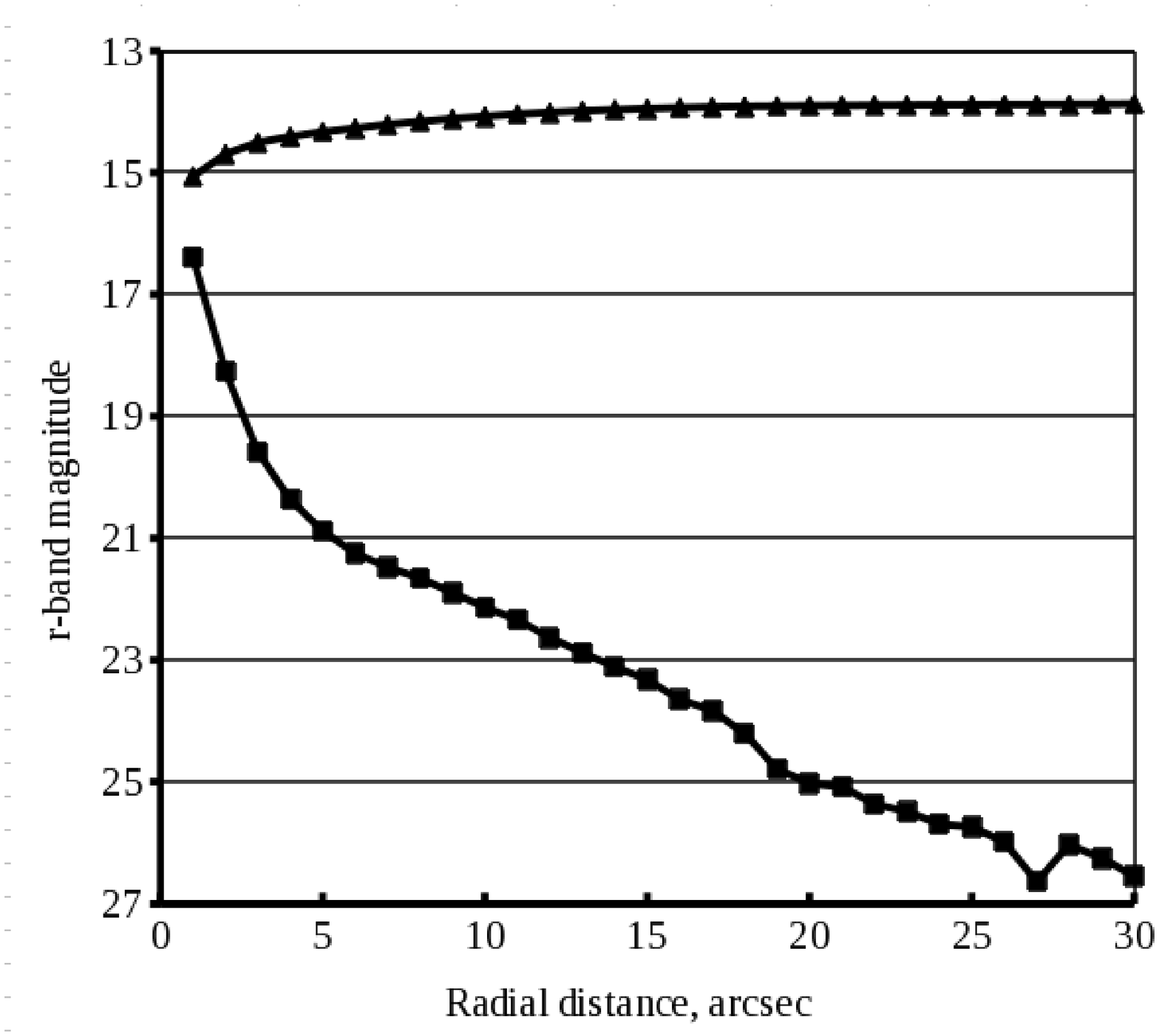,width = .85\linewidth}
\end{minipage}
\hfill
\begin{minipage}[t]{.45\linewidth}
\centering
\epsfig{file = 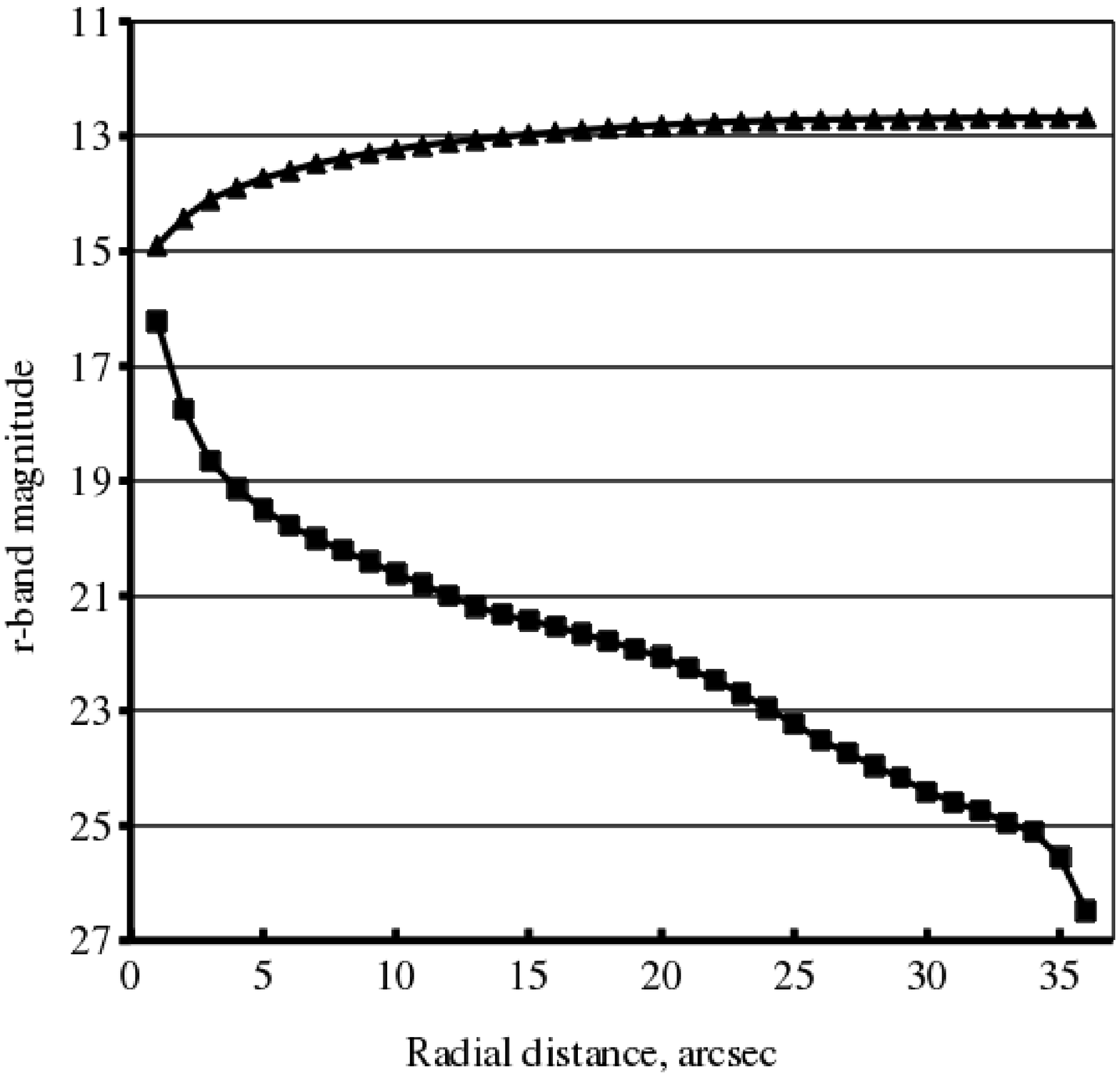,width = .85\linewidth}
\end{minipage}
\caption{Brightness profiles of Mrk 704 (left) and Mrk 766 (right). Lower graphs - surface brightness profiles, magnitude from square arcsecond. Upper graphs - magnitude in circle of corresponding radius.}\label{fig1}
\end{figure}

\begin{table}
 \centering
 \caption{Parameters of galaxies. r - r-band SDSS magnitude; $a$ - SDSS r-band major semiaxis in arcseconds; $a_{back}$ - outer radius of background region; $m_c$ - averaged surface brightness in central circle with radius 1$''$, in $^m/\square ''$; $m_a$ - surface brightness at radial distance $a$; $m_1$, $m_2$ and $m_3$ - estimated magnitudes within circles of radius 1, 2 and 3 arcseconds correspondingly. Notes: $^*$ For 3 galaxies there were no SDSS magnitudes and diameters. We used 2MASS K magnitudes in largest raduis from available: for NGC 3227 - 54.2$''$, for NGC 4051 - 35.4$''$ and for Ark 120 - 25$''$; $^{**}$ For these galaxies we used large semiaxis from isophote K=$ 20^m/\square ''$ }\label{tab1}
 \vspace*{1ex}
 \begin{tabular}{|c|c|c|c|c|c|c|c|c|}
  \hline
Name & r & $a"$ & $a_{back}"$ & $m_c$ & $m_a$ & $m_1$ & $m_2$ & $m_3$ \\
\hline
NGC 3227 & 11.30$^*$ & 92.6$^{**}$ & 120 & 15.41 & $>$26.5 & 14.10 & 13.19 & 12.78 \\
\hline
NGC 4151 & 11.63 & 102.60 & 120 & 16.98 & $>$26.5 & 15.66 & 14.47 & 13.64 \\
\hline
NGC 4051 & 12.03$^*$ & 102.60$^{**}$ & 120 & 16.04 & $>$26.5 & 14.73 & 14.12 & 13.83 \\
\hline
NGC 5548 & 12.63 & 39.43 & 49 & 16.05 & $>$26.5 & 14.74 & 14.15 & 13.76 \\
\hline
Mrk 590 & 12.87 & 37.17 & 56 & 16.34 & $>$26.5 & 15.02 & 14.29 & 13.86 \\
\hline
NGC 7603 & 12.96 & 50.37 & 60 & 16.98 & 24.99 & 15.67 & 14.70 & 14.25 \\
\hline
Ark 120 & 12.96$^*$ & 16.10$^{**}$ & 29 & 15.28 & $>$26.5 & 13.97 & 13.34 & 13.12 \\
\hline
Mrk 766 & 13.01 & 35.43 & 120 & 16.22 & 25.56 & 14.90 & 14.42 & 14.10 \\
\hline
Mrk 79 & 13.38 & 45.38 & 120 & 17.66 & 25.25 & 16.34 & 15.40 & 14.87 \\
\hline
Mrk 817 & 13.66 & 23.45 & 80 & 16.47 & 26.07 & 15.15 & 14.53 & 14.26 \\
\hline
Mrk 704 & 14.04 & 29.09 & 112 & 16.38 & 26.25 & 15.07 & 14.70 & 14.51 \\
\hline
Mrk 382 & 14.66 & 23.95 & 120 & 17.58 & 25.52 & 16.26 & 15.57 & 15.28 \\
\hline
Mrk 290 & 14.72 & 15.42 & 120 & 16.86 & 24.70 & 15.54 & 15.12 & 14.94 \\
\hline
Mrk 504 & 14.87 & 20.77 & 76 & 17.63 & 25.20 & 16.31 & 15.78 & 15.46 \\
\hline
Mrk 110 & 15.21 & 25.50 & 104 & 18.19 & 25.60 & 16.88 & 16.32 & 16.05 \\
\hline
3C 332 & 16.78 & 8.23 & 29 & 18.71 & 25.19 & 17.39 & 17.00 & 16.78 \\

 \hline 
 \end{tabular}

\end{table}

\section*{Results and conclusions}
\indent \indent Examples of surface brightness profiles are shown at Fig. 1. Profiles of disk component up to $ 25^m/\square ''$ are close to linear and active nuclei are more bright. Numerical results are presented in Table 1. In smaller galaxies at radial distance corresponding to major semiaxis we see surface brightness close to $ 25^m/\square ''$. This indicates that performed calculations were correct. For larger galaxies we obtained $ 25^m/\square ''$ at less radial distances. One of possible sources of diversity may be uncertainity of diameter values. At r-band SDSS images background level is 1110$\pm$7 counts per pixel (pixel size is 0.4$''$, exposition is 54 sec) that corresponds to $ 18.6^m/\square ''$. With such a large background level, a source of 25$^m$ can be detected only when observing of the sky area of at least 20 $\square''$. So error of isophotal diameter can not be less than 5$''$. Note that r-band SDSS diameters in NED are given with 0.01$''$ accuracy and without errors. For the obtaining of AGN light curves the emission of only the central part of the galaxy should be considered. So we have not performed accurate determination of brightness profiles of outer parts of galaxies. The main result of our work is magnitudes of central part of galaxy image with different radius. This brightness should be subtracted from concrete variability observation to find the brightness of AGN without the buldge. Taking into account the notes mentioned above we can conclude that obtained profiles can be useful for research of variability of Seyfert galaxies. 

\section*{Acknowledgement}
\indent \indent We would like to thank V.T.Doroshenko for suggesting the problem.

\end{document}